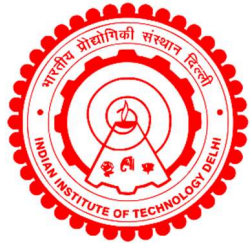

Indian Institute of Technology, Delhi

ELL893: Cyber Physical Systems

*Final Project Report*

*on*

**"Smart Home"**

**Muneeb Ahmed** (2020BSZ8703), **Mohd Majid Akhtar** (2021BSZ8184)

10[th] November 2021

**Abstract**

This project discloses the development of a solution for realizing a smart home in the post COVID-19 era using the Internet of Things domain knowledge. COVID-19 outbreak has been catastrophic and impacted everyone's lives due to its rapid transmission from one body to another. This project aims to reduce virus transmission by eliminating the need to touch any common-point surface in a home, such as switches, doorknobs, and remotes. We provide a generic solution by coupling things with the internet to control them remotely. The project aims to showcase a working solution to controlling devices like smart bulbs, smart fans, smart ACs, and smart door locks in our self-developed emulator over WWW securely using two different protocols, viz., HTTP and MQTT-over-WSS. Additionally, intent authentication over HTTP is based on digital signature that is demonstrated using RSA (encryption) and MD5 (hashing) when the system is deployed in insecure environment(s). RESTful API deployed on AWS EC2/Lambda is used to realize HTTP communication protocol, and MQTT is realized using AWS IoT service. The developed project can be applied to any smart home setting like a hotel or public place using AWS IoT, Lambda, or similar infrastructure as a broker.

## I. INTRODUCTION

Before the COVID-19 outbreak, smart homes were perceived as a luxury and not a necessity. However, the need for owning devices that can be remotely accessed, monitored, and controlled is increasing day by day. We assume this to be one of the motivating factors behind the predicted score of 30.9 billion devices by the end of 2025, an incredible growth according to the Statista report [1]. COVID-19 transmission has been dramatically influenced by touching surfaces of common points in the home or outside, like doorknobs, switches, and remote of TV and AC [2]. It continues to pose challenges to the healthcare system to its undue nature. Hence, this project aims to provide innovative solutions within the realms of technology to reduce the need to touch surfaces of common areas. With the advancement in the information and communications technology (ICT) sector, we can now achieve the required goal.

This project attempts to mimic the smart home ecosystem using the Internet of Things (IoT) domain knowledge for building a home automation system. Given the security issues of publicly transmitted data over the internet, it is necessary to design an authenticated system to fulfill the required need. To accomplish the stated problem, we have developed a technology stack as a set of modules, viz.,

1. **IoT Emulator App** (*that features various IoT devices*)
2. **Device App** (*that acts as a broker for pub-sub instances*)
3. **Client App** or **UI** (*by which users can control the state of IoT devices*)
4. **Key Distribution Center App** (*used only in case of HTTP*)

We have designed our own simulation environment (Emulator App), an HTML web page consisting of four kinds of smart devices: *bulb*, *fan*, *AC*, and *door lock*. The Client App features authenticated front-end dashboard where the user interacts with the devices. The core of the system is the Device App (responsible for behavior mapping). The KDC App (used in HTTP protocol) is a trusted party that acts as a public-key provider/verifier for the intent of the Client App. When an authenticated user publishes its intent to change the state of the devices, the Client App sends RESTful API (or MQTT) requests to the Device App to process the intent. The Device App is responsible for controlling the behavior of the intent(s) it receives. Device App is deployed on HTTP using AWS EC2, while on MQTT it is served through AWS IoT. The choice of the protocol (HTTP or MQTT) is offered as a will to the user. In both cases, the end-to-end system has been realized in real-time. Most of the deliverables for this system have been completed, some of which are as follows:

1) **Security/Integrity**
   a. Credentials based authentication control (for Client App)
   b. Digital signature based intent verification (for security on HTTP protocol)
   c. Encryption + Authentication (for security on MQTT protocol)
2) **Logs and Reports** (for logging events)
3) **Rules/Security Policie**s (for defining the correct behavior)
4) **Performance Analysis of two communication protocols**: (HTTP and MQTT over WebSocket)
5) **Real-time implementation** (end-to-end demonstration for security, authentication, and control)

II.       PROBLEM STATEMENT

The world has observed the unprecedented COVID-19 outbreak, and it impacted the lives of millions. On the other hand, ICT or Industrial 4.0 has fully immersed into reality during this time. The transmission of COVID spread rapidly due to people being in proximity and sharing surfaces of everyday objects like doorknobs, switches, and various appliances' remote. Therefore, it is essential to implement the technology to help reduce transmission and increase more contact-free and contact-less operations of daily life?

Another very inconspicuous issue with public domain transmission of data lies within the scope of **providing secure and authenticated access to the devices**. The control of critical devices can be disastrous if security challenges are not handled in priority. The processing power on such devices is assumed to be limited. Hence, the communication control should be lightweight, in addition to, being secure (also referred to as *lightweight-key security*).

**Timeliness** is another core issue while deploying such systems. The control of devices in stochastic environments must be ensured to be fast. While researchers around the world work towards improvements in decentralized systems (such as Edge or Fog Computing) that tend to minimize bandwidth and latency utilization on network-level (or physical level), it is also essential for application developers to provide timely message passing on implementation level (or application level).

While designing such systems, the **entities must be autonomous and scalable.** They must interact with each other in isolation. It must be possible to create multiple instances of each entity (such as Client App, Device App or Emulator App). It is essential for any scalable system like this to perform tasks in composition (when implemented in distributed environments). The design-level issues of such systems include **unavailable or expensive simulation environments**. Such scalable software/hardware are important for research/implementation in controlled environments before deployment in real-world.

We intend to implement security-aware end-to-end API-based control to demonstrate this work, featuring two protocols, viz., MQTT, and HTTP. API based mechanism will allow remote authenticated access to control the system over WWW. **Our generalized system is autonomous on entity level of the design.** All its autonomous entities have the potential to be deployed at the network edge. We have designed a custom Emulator App to simulate results in a controlled environment as shown in Figure 1.

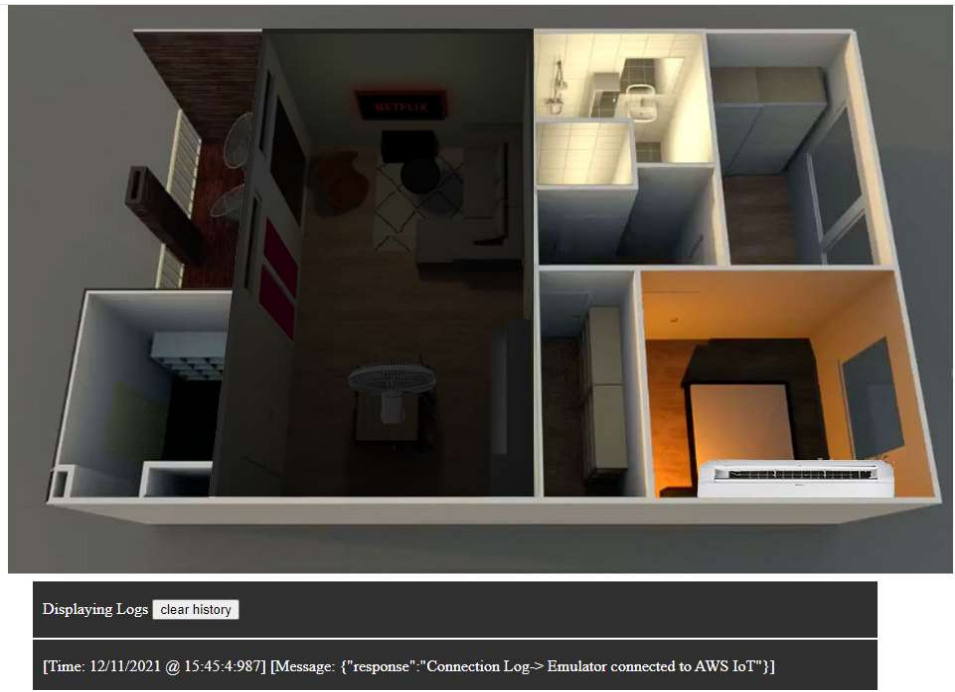

Figure 1. Emulator running as a webpage

### III. PROPOSED SOLUTION

This section have is divided into three sub-sections: Materials and methods, Implementation (in HTTP and MQTT), and API flow.

#### A. Materials and Methods

We propose an architecture that utilizes the following set of technologies:

1) **Emulator App:** HTML, CSS and jQuery, for building the environment using a top view image of a 3D model of a home.
2) **Client App:** Credential-authenticated WebApp for calling the defined set of APIs. The commands also include *composite commands*. This composes the intent packet that contains control information for each individual device.

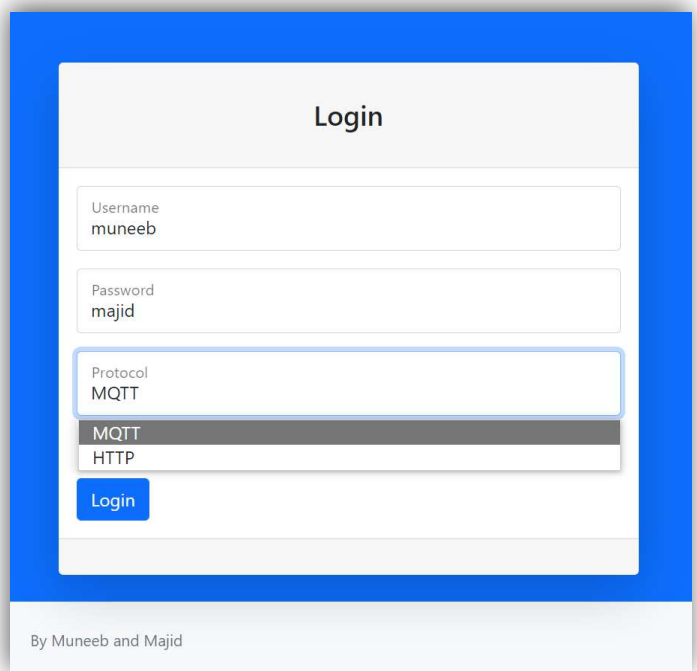

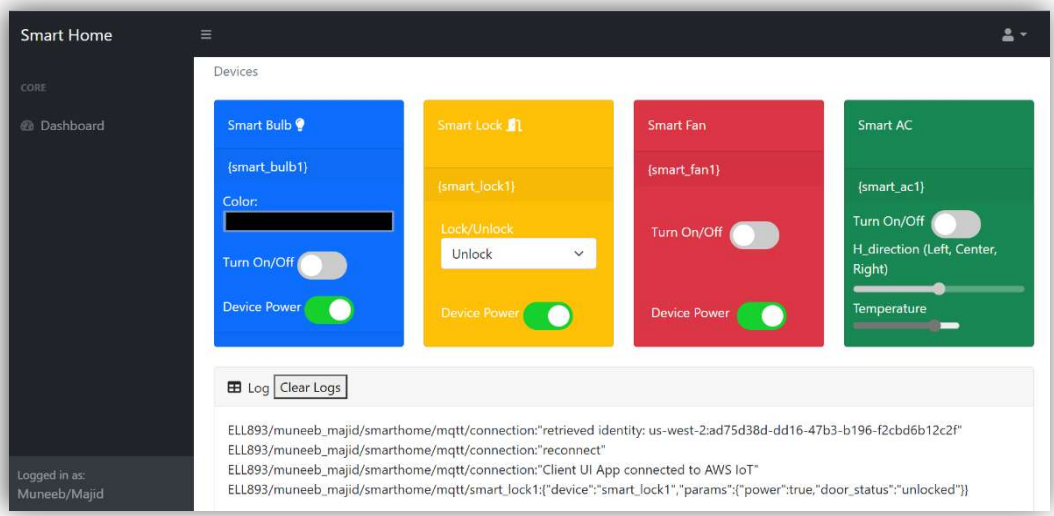

Figure 2. Showing Client (UI) App

3) **Device App (MQTT):** Deployed with browserified AWS IoT core JavaScript SDK [3] to subscriber/publisher topics.
4) **Device App (HTTP):** Python based RESTful API deployed on AWS EC2 [4] for creation and policy mapping to publisher/subscriber interfaces.
5) **KDC App:** RESTful API (*for HTTP protocol only*) for creating/verifying public key of intent-provider (i.e., client).

B. **Implementation (HTTP protocol)**

**Interfaces/Entities in API Communication**

There are three bidirectional interfaces deployed in the end-to-end API follow of this work:
1. **Interface 1:** Between UI (Client App) and Device App
2. **Interface 2:** Between Device App and Emulator App
3. **Interface 3:** Between any party and KDC App

**{apiRoot}** signifies endpoint URI. For AWS EC2, it has been set to http://54.212.46.254:5000.
**On*Entities are shown in figure below.***

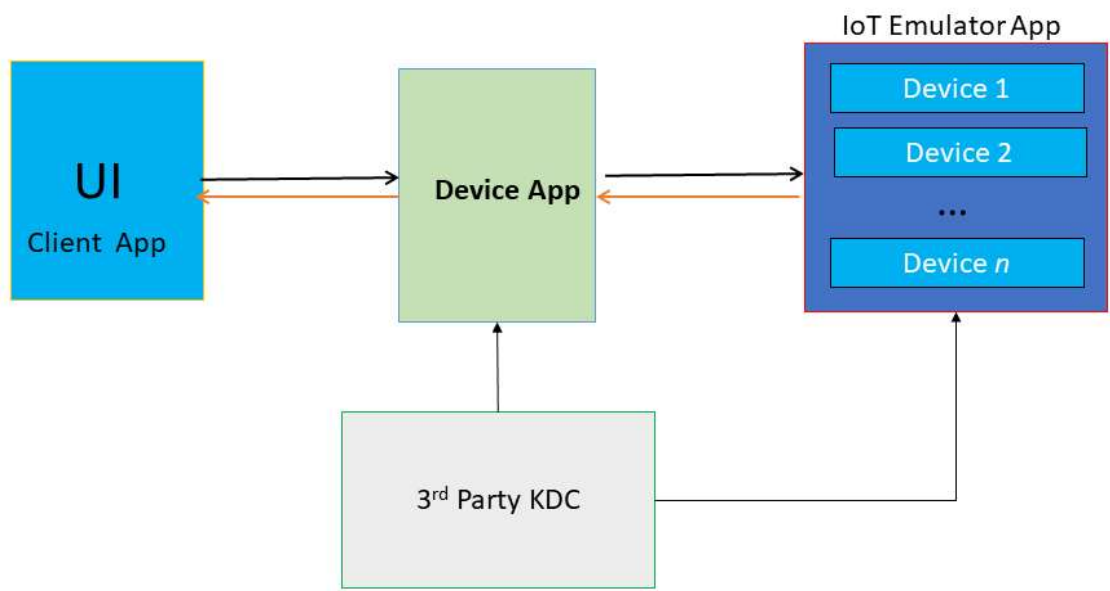

Figure 3. Entities in HTTP API Communication

### C. HTTP API FLOW
### 1. Between Client App and Emulator through Device App

This has been divided into two functions:

**a)** The Client App can send POST request to Device App along with devId (to which it wants to command) and the command itself (called intent, that contains status of parameters for the device). MD5 or any other Hash of the intent packet must be encrypted by the Client App and should be appended along with the packet (this process is digitally signing the packet). The packet is verified by the Device App using the public key of the Client App which it receives from KDC. Device App holds the current status of the devices after verifying the digital signature of the client from KDC. Emulator periodically sends a GET request to Device App to know the current state of Devices.

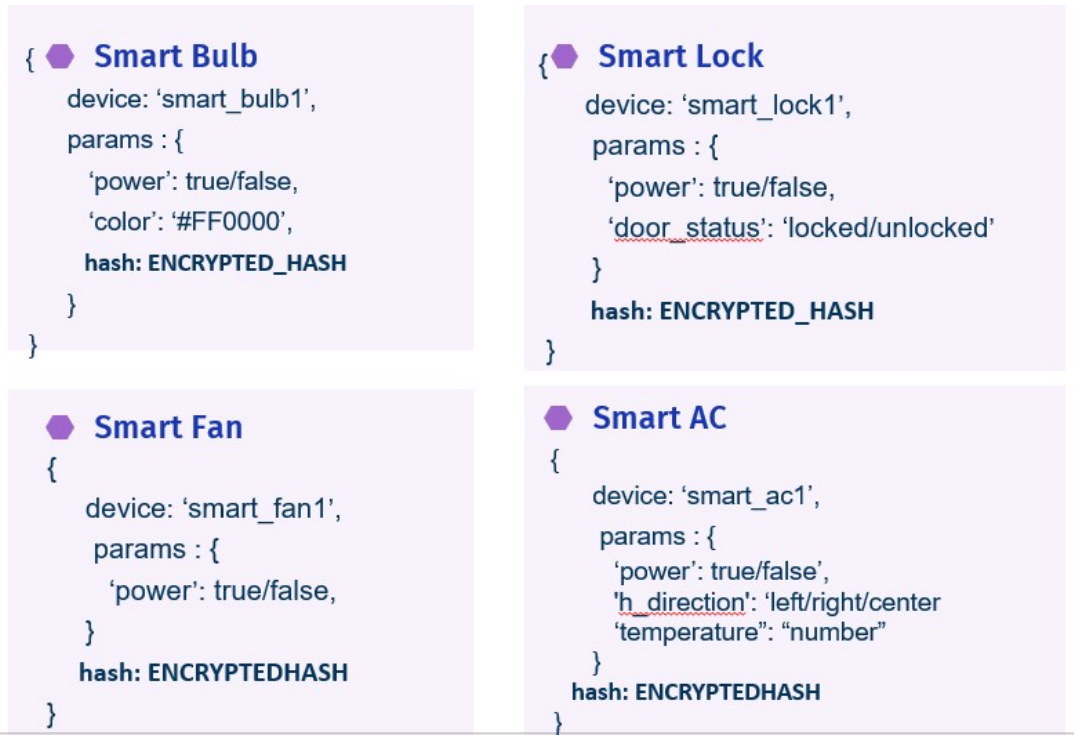

Figure 4: Packet format in HTTP

b) The Client App can send GET request to Device App to know the status of online devices.

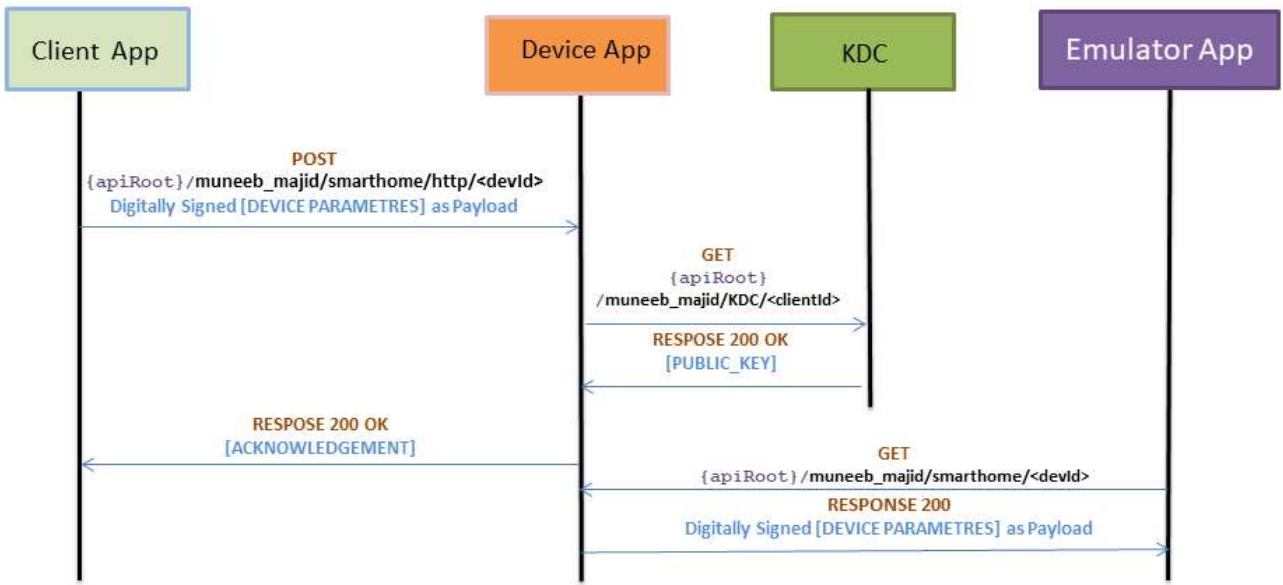

Figure 5. API Flow of communication

## 2. Between Device/Emulator App and KDC App

Initially, KDC acts as a repository of public keys of Client App and Emulator. The distribution of private keys is assumed to be done apriori. This interface handles providing public key of Client App for verification of digital signature.

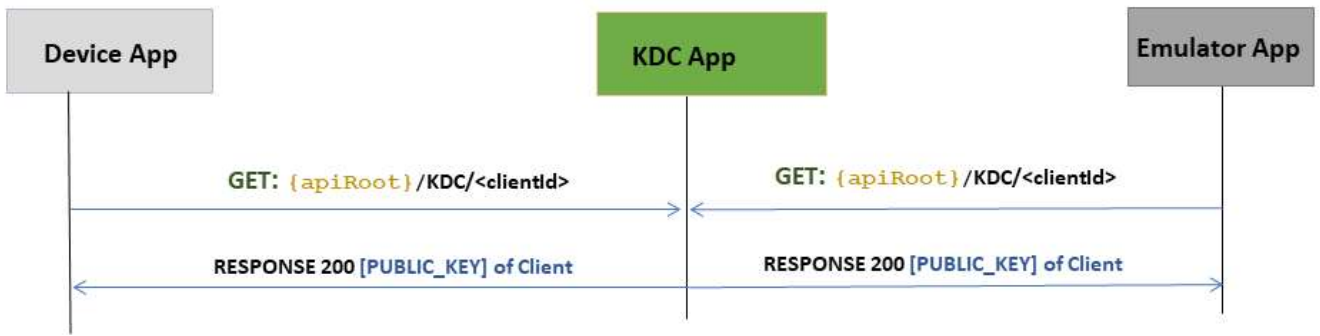

Figure 6. Flow of Keys

### D. Implementation (MQTT over WebSocket protocol)

The steps in which we have designed the smart home application using **MQTT over WSS communication protocol**:

1. Since our Emulator is browser-based. Hence, we have used AWS IoT JavaScript SDK to interact with the AWS IoT Core platform.
2. Due to browser-dependency, we cannot use the build JavaScript SDK directly. Hence, we have used Browserify to convert the SDK into a browser-supported client side JavaScript as a *aws-iot-sdk-browser-bundle.js* file.
3. To use above generated file, we first added a 'Thing' as 'Emulator' to the AWS IoT Core to receive *AWSIoTEndpoint*.
4. Then, we have added an Identity in the AWS Cognito Identity Pool Console to allow identities to use AWS services using the secret *CognitoIdentityPoolID*.
5. We have then added a policy in AWS IAM to use IoT resources.

The *high level flow* diagram is better analysed in the figure 7 below:

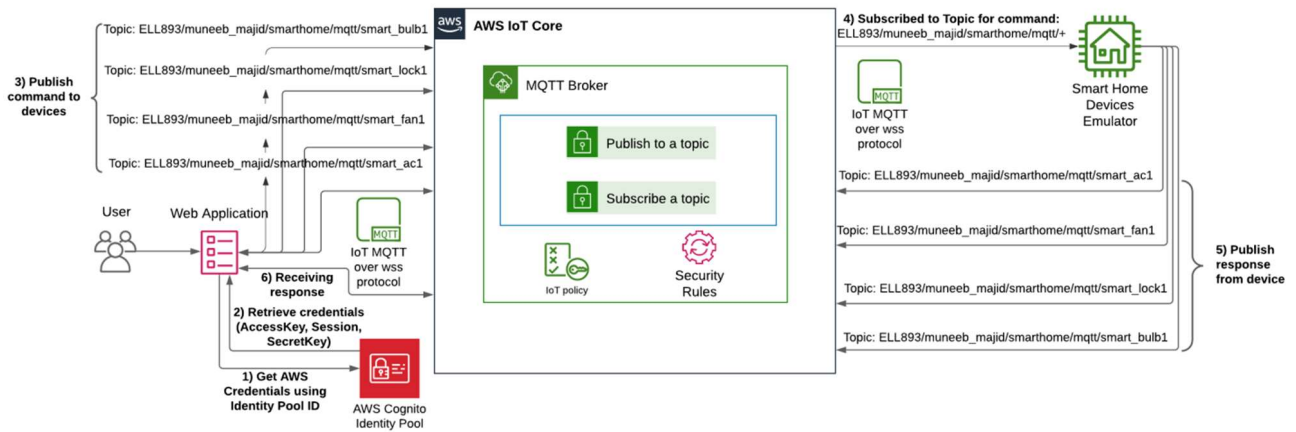

Figure 7. Smart Home Architectural Diagram

The *low level* implementation details are as follow:

1) First we are handling the payload to be sent on the AWS IoT to a specific topic from the Client UI app in JSON format as shown in table below.

| Table 1: MQTT over WebSocket Payload for All Four Devices | | |
|---|---|---|
| **Device Type** | **Payload Content for command** | **Information in Params** |
| Smart Bulb | `{`<br>    `"device": "smart_bulb1",`<br>    `"params":{`<br>        `"power":true,`<br>        `"color":"#ffffff"`<br>    `}`<br>`}` | //can be true or false<br>//can be hex code to change color |
| Smart Lock | `{`<br>    `"device": "smart_lock1",`<br>    `"params":{`<br>        `"door_status":"locked"`<br>    `}`<br>`}` | //can be locked or unlocked |
| Smart Fan | `{`<br>    `"device": "smart_fan1",`<br>    `"params":{`<br>        `"power":true`<br>    `}`<br>`}` | //can be true or false |
| Smart AC | `{`<br>  `"device": "smart_ac1",`<br>    `"params":{`<br>        `"power":true,`<br>        `"h_direction":"rotate(0deg)",`<br>        `"temperature":20`<br>    `}`<br>`}` | //'power' can be true or false<br>//'h_direction' can be rotate(0deg) for center, rotate(-45deg) for left and rotate(45deg) for right.<br>//can be in range 18-26 |

2) Then we defined the Policy in AWS IAM:

The policy that we have written in IAM to use AWS IoT Core services while authenticating using AWS Cognito Identity Pool is shown below:

```
{
    "Version": "2012-10-17",
    "Statement": [{
        "Effect": "Allow",
        "Action": [
            "iot:*"
        ],
```

```
            "Resource": "*"
    }]
}
```

However, while in production environment, we would recommend using a fine-grained restricted action and resources properties. This is done by defining a restricted policy to limit users and devices to publish and subscribe data only to the specified topic and not to any other resource as shown below.

```
{
        "Version": "2012-10-17",
        "Statement": [{
           "Effect": "Allow",
           "Action": [
                 "iot:Publish"
           ],
           "Resource": [
              "arn:aws:iot:us-west-2:974628150977:topic/ELL893/muneeb_majid/smarthome/mqtt/*"
           ]
        },
        {
           "Effect": "Allow",
           "Action": [
                 "iot:Subscribe"
           ],
           "Resource": [
              "arn:aws:iot:us-west-2:974628150977:topicfilter/ELL893/muneeb_majid/smarthome/mqtt/*"
           ]
        }
    ]
}
```

3) The overall functionality:

The client app is under a repository '/client' where it uses *index.html* as the main webpage and the UI of the APP. The JavaScript function calls are written in file *bundle.js* where each functions are called with onChange() function from the respective HTML element. The logs are provided in the UI with different color codes representing error logs, connection logs, command logs and response logs. We have used four topics respectively for publishing commands to Emulator for different devices as shown in above table. We are also receiving back response from the Emulator in the same respective device topic. We also have one connection topic where Client UI and Emulator are sending message to each that they are connected to the AWS IoT.

Similarly, the emulator app is under a repository '/emulator' where it uses *index.html* as the main webpage and the emulator. The JavaScript function calls are written in file *emulator_index.js* where it is performing changes respective to the HTML element according to the command received from the client app. The logs are provided in the emulator for showing command received and the response sent.

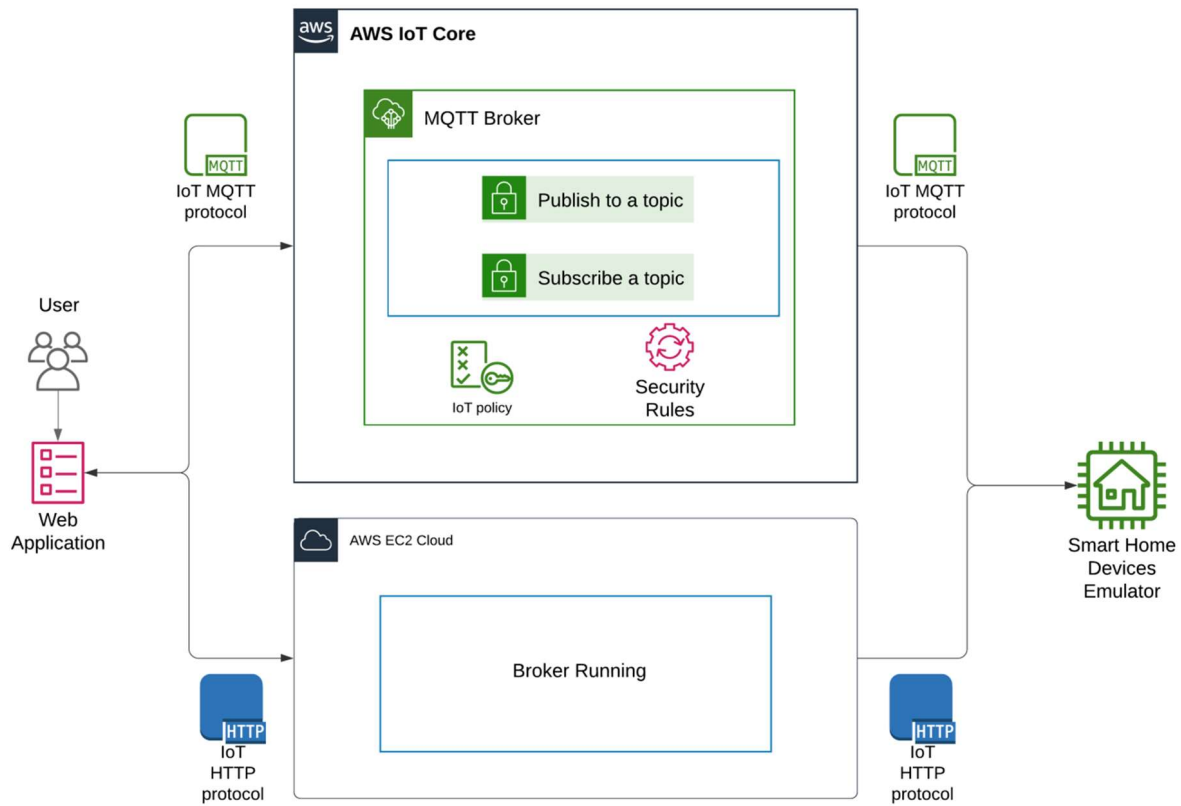

Figure 8. Smart Home Overall Architectural Diagram

## IV. EVALUATION AND DELIVARABLES

*The deliverables of this project are as follows:*

1. Implementing real-time Smart Home control on two protocols (HTTP and MQTT).
2. Simulate the results in emulator.
3. Perform time analysis on two protocols (HTTP and MQTT).
4. Comparatively discuss the issues with implementation/protocols.
5. Implement authentication, security policies, and digital signature-based security/integrity on messages.

We have evaluated our system on the following set of parameters viz., *Evaluation of Protocols, Evaluation of adding digital signature verification, and Evaluation of overall security of the system.*

### A. Evaluation of Protocols:
HTTP v/s MQTT over WSS

1. MQTT is light weight in comparison to HTTP considering payload size and method calls.
2. HTTP is request-response protocol for client-server computing and MQTT is data centric which transfers data as byte array.
3. Latency and bandwidth usage are optimal in MQTT but not optimized in HTTP.

## B. Evaluation of adding digital signature verification

In HTTP environment, we had to add security features manually by adding hash of the message and encrypting the hash with the client's private key. This hash was appended to the message and sent to the Device App. There is an intrinsic delay of long keys are used. We have used MD5 hashing with RSA (1048-bit key) which promises lightweight and quick computation. Yet this adds an infinitesimal unavoidable delay of the order of microseconds. For verification, the Device App sends request to KDC to receive a public-key and decrypt the message. There is a minute delay in making this request. All these delays have been mentioned in following table.

Table 2: Delay for different tasks

| STAGE | Delay |
|---|---|
| Digitally Signing the packet | 20 ms |
| Sending Data to Device App | 480 ms |
| Verification of Digital Signature (including public key fetch from KDC) | 525 ms |
| Receiving Data at the Emulator App | 426 ms |
| **TOTAL DELAY** | 1451 ms |

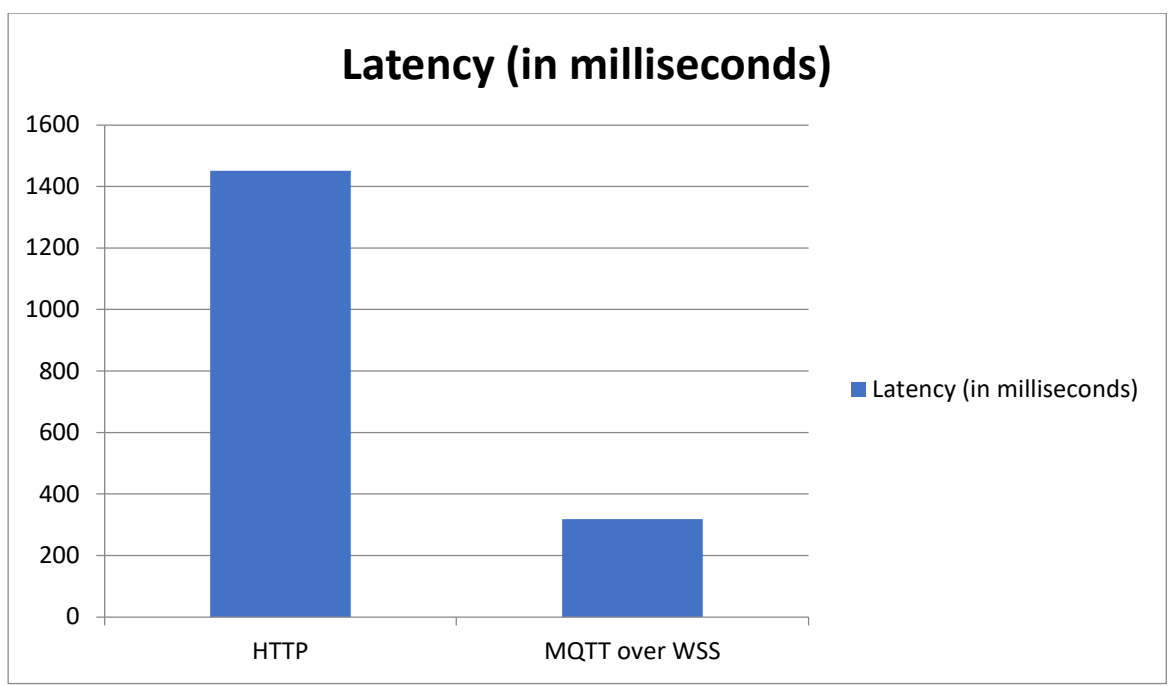

Figure 11: Latency comparison between MQTT over WSS and HTTP

## C. Evaluation of overall security

MQTT-over-WSS is offered to be serviced by AWS in-house IoT service, the security of which is collaborated into AWS security policies itself. It provides secure transaction of messages on individual device level as well as emulator level. The security on HTTP is provided additionally as it does not have its own authentication mechanism against man in the middle attack (unless HTTPs is used). Upon adding digital signature scheme, the implemented system is robust to a variety of attacks.

## D. Logs and Reports

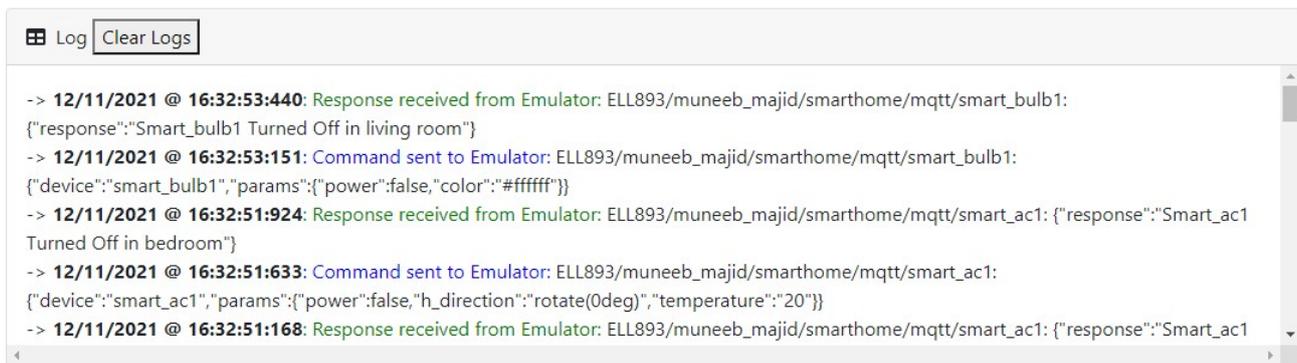

Figure 10: Logs shown in the UI

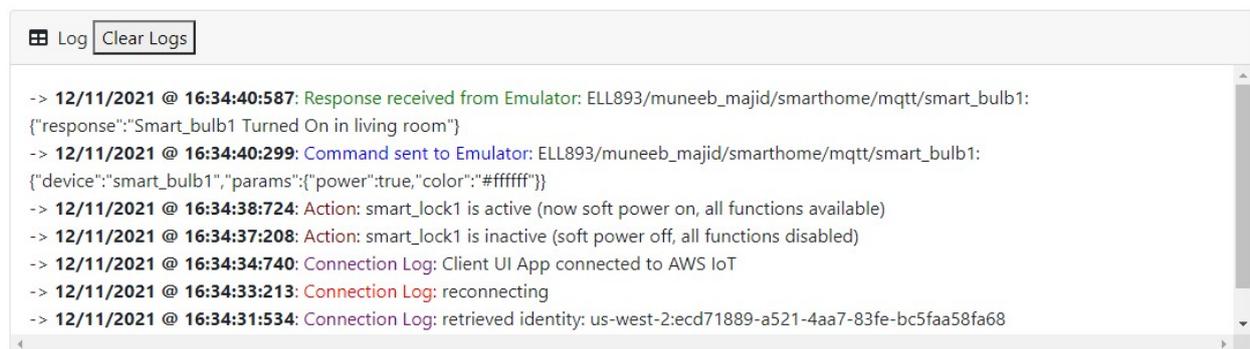

Figure 11: More Logs shown in the UI with different color codes

We envision our solution is beneficial towards the society in reducing the transmission of of novel COVID. Along with it, the developed project can be applied to any smart home setting like a hotel or public place where social distancing is required to reduce the transmission of virus by touching surfaces.

**Source Code: (For both emulator and client app using different communication protocols)**

1) MQTT over WebSocket using AWS IoT: https://github.com/mohdmajidakhtar/ELL893
2) HTTP using AWS EC2: https://muneebpandith.github.io/ELL893

**References**

1. IoT and non-IoT connections worldwide 2010-2025, Statista Report, 2021. (Online)- https://www.statista.com/statistics/1101442/iot-number-of-connected-devices-worldwide/
2. Coronavirus disease (COVID-19): How is it transmitted?, WHO, 2021. (Online)- https://www.who.int/news-room/q-a-detail/coronavirus-disease-covid-19-how-is-it-transmitted
3. Amazon IoT Core, https://aws.amazon.com/iot-core/
4. AWS EC2, https://aws.amazon.com/ec2/